# ROBustness In Network (robin): an R package for Comparison and Validation of communities

*by Valeria Policastro, Dario Righelli, Annamaria Carissimo, Luisa Cutillo and Italia De Feis*

**Abstract** In network analysis, many community detection algorithms have been developed, however, their implementation leaves unaddressed the question of the statistical validation of the results. Here we present **robin** (ROBustness In Network), an R package to assess the robustness of the community structure of a network found by one or more methods to give indications about their reliability. The procedure initially detects if the community structure found by a set of algorithms is statistically significant and then compares two selected detection algorithms on the same graph to choose the one that better fits the network of interest.

We demonstrate the use of our package on the American College Football benchmark dataset.

## Introduction

Over the last twenty years, network science has become a strategic field of research thanks to the strong development of high-performance computing technologies. The activity and interaction of thousands of elements can now be measured simultaneously allowing us to model cellular networks, social networks, communication networks, power grids, trade networks, just to cite a few examples. Different types of data will produce different types of networks in terms of structure, connectivity and complexity. In the study of complex networks, a network is said to have community structure, if the nodes are densely connected within groups but sparsely connected between them (Girvan and Newman, 2002). The inference of the community structure of a network is an important task. Communities allow us to create a large-scale map of a network since individual communities act like meta-nodes in the network, which makes its study easier. Moreover, community detection can predict missing links and identify false links in the network. Despite its difficulty, a huge number of methods for community detection have been developed to deal with different size complexity and made available to the scientific community by open-source software packages. In this paper we will address a specific question: are the detected communities significant or are they a result of chance only due to the positions of edges in the network?

An important answer to this question is the Order Statistics Local Optimisation Method (*OSLOM*, http://www.oslom.org/) presented in Lancichinetti et al. (2011). *OSLOM* introduces an iterative technique based on the local optimisation of a fitness function, the C-score (Lancichinetti et al., 2010), expressing the statistical significance of a cluster with respect to random fluctuations. The significance is evaluated fixing a threshold parameter P a priori.

Another interesting approach is the Extraction of Statistically Significant Communities (*ESSC*, https://github.com/jdwilson4/ESSC) technique proposed in Wilson et al. (2014). The algorithm is iterative and identifies statistically stable communities measuring the significance of connections between a single vertex and a set of vertices in undirected networks under the configuration model (Bender and Canfield, 1978) used as the null hypothesis. The method employs multiple testing and false discovery rate control to update the candidate community.

Kojaku and Masuda (2018) introduced the QStest (https://github.com/skojaku/qstest/) a method to statistically test the significance of individual communities in a given network. Their algorithm works with different detection algorithms using a quality function that is consistent with the one used in community detection and takes into account the dependence of the quality function value on the community size. QStest assesses the statistical significance under the configuration model too.

Very recently, He et al. (2020) suggested the Detecting statistically Significant Communities (DSC) method, a significance-based community detection algorithm, that uses a tight upper bound on the p-value under the configuration model coupled with an iterative local search method.

OSLOM, ESSC and DSC assess the statistical significance of every single community analytically while QStest adopts the sampling method to calculate the p-value of a given community. Moreover, all of them detect statistically significant communities under the configuration model, and only QStest is independent of the detection algorithm.

We present **robin** (ROBustness In Network) an R/CRAN package whose purpose is to give clear indications about the reliability of one or more community detection algorithms under study analyzing its robustness with respect to random perturbations. The idea behind **robin** is that if a partition is significant, it will be recovered even if the structure of the graph is modified while if the partition is not significant minimal modifications of the graph will be sufficient to change it. **robin** gets inspired by the concept presented by Carissimo et al. (2018), who studied the stability of the recovered partition against random perturbations of the original graph structure using tools from Functional Data Analysis (FDA).

**robin** provides the best choice among the variety of the existing methods for the network of interest. It is based on a procedure that gives the opportunity to use the community detection techniques implemented in igraph package Csardi and Nepusz (2019), while providing the user with the possibility to include other community detection algorithms. **robin** initially detects if the community structure found by some algorithms is statistically significant, then it compares the different selected detection algorithms on the same network. **robin** assumes undirected graphs without loops and multiple edges.

**robin** looks at global stability of the detected partition and not of single communities, but accepts any detection algorithm and any random model, and these aspects differentiate it from OSLOM, ESSC, DSC and the QStest. Moreover, unlike other studies that treat the comparison between algorithms in a theoretical way such as Yang et al. (2016), **robin** aims to give a practical answer to such comparison that can vary with the network of interest.

## The Model

**robin** implements a methodology that examines the stability of the recovered partition by one or more algorithms. The methodology is useful for two purposes: to detect if the community structure found is statistically significant or is a result of chance and to choose the detection algorithm that better fits the network under study. These are implemented following two different workflows.

The first workflow tests the stability of the partitions found by a single community detection algorithm against random perturbations of the original graph structure. To address this issue we specify a perturbation strategy (see subsection **Perturbation strategy**) and a null model to build some procedures based on a prefixed stability measure (see subsection **Stability measure**). Given:

- a network of interest $g1$
- its corresponding null random model $g2$
- a Detection Algorithm (DA)
- a stability measure (M)

our process builds two curves as functions of the perturbation level $p$ as shown in Figure 1 and tests their similarity by two types of functional statistical tests (see subsection **Statistical Tests**).

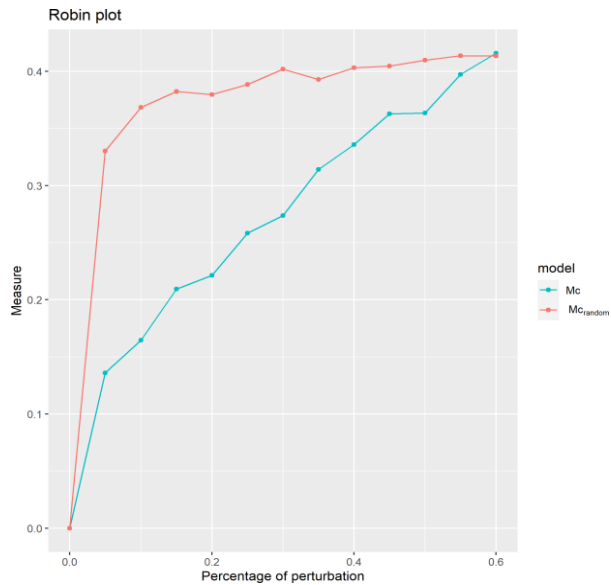

**Figure 1:** Example of $Mc_{random}$ and $Mc$ curves generated by an M stability measure

The first curve $Mc$ is obtained computing M between the partition of the original network $g1$ and the partition of different perturbed version of $g1$. The second curve $Mc_{random}$ is obtained by computing

M between the partition of a null random network $g2$ and the partition of different perturbed version of $g2$.

The comparison between the two M curves enables us to reconsider the problem regarding the significance of the retrieved community structure in the context of stability/robustness of the recovered partition against perturbations. The basic idea is that if small changes in the network cause a completely different grouping of the data, the detected communities are not reliable. For a better understanding of this point, we suggest the reader refer to the original paper Carissimo et al. (2018) where the methodology was developed.

The choice of the null model plays a key role because we would expect it to reproduce the same structure of the real network but with completely random edges. For this reason, **robin** offers two possibilities: a degree preserving randomization by using the `rewire` function of the **igraph** package or a model chosen by the user.

The degree preserving randomization, i.e. Configuration Model (CM), is a model able to capture and preserve strongly heterogeneous degree distributions often encountered in real network data sets and is the standard null model for empirical patterns. Nevertheless, it can happen that it is not sufficient to preserve only the degree of the graph understudy, so **robin** allows the user to include their own null model.

In section **Example test: the American College football network** we explore the *dk* null random model provided in Orsini et al. (2015), whose code is available at https://github.com/polcolomer/RandNetGen as a possible alternative to CM. The *dk*-series model generates a random graph preserving the global organization of the original network at various increasing levels of details chosen by the user via the setting of the parameter *d*. More precisely, the *dk*-series is a converging series of properties that characterize the local network structure at an increasing level of detail and define a corresponding series of null models or random graph ensembles. Increasing values of *d* capture progressively more properties of the network: *dk* 1 is equivalent to randomize the network fixing only the degree sequence, *dk* 2 fixes additionally the degree correlations, *dk* 2.1 fixes also the clustering coefficient and *dk* 2.5 the full clustering spectrum.

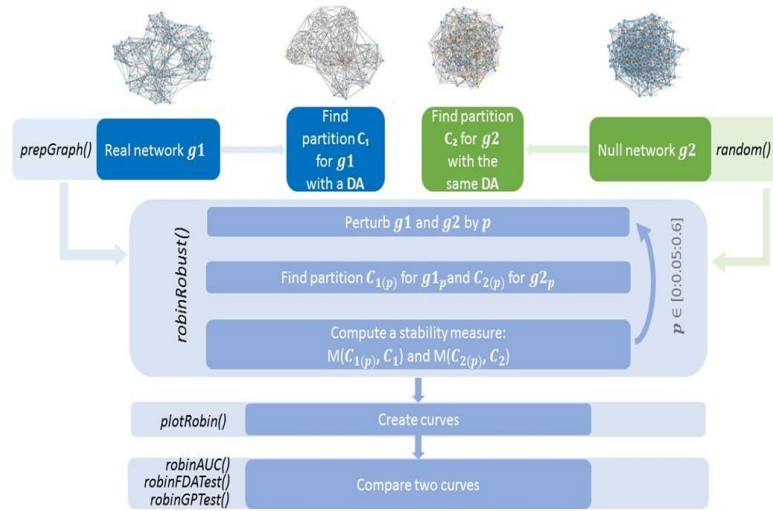

**Figure 2:** Flowchart summarizing the procedure for testing the goodness of a community detection algorithm.

The first workflow is summarised as follows:

1. finds a partition $C_1$ for the real network and a partition $C_2$ for the null network,
2. perturbs both networks,
3. retrieves two new partitions $C_{1(p)}$ and $C_{2(p)}$,
4. calculates two clustering distances (for the real network and the null network) between the

original partitions and the ones obtained from the perturbed network as:

$$M\left(C_{1(p)}, C_1\right) \quad \text{and} \quad M\left(C_{2(p)}, C_2\right) \qquad (1)$$

Steps 2) - 4) are computed at different perturbation levels $p \in [0 : 0.05 : 0.6]$ to create two curves, one for the real network and one for the null model, then their similarity is tested by two functional statistical tests described in subsection **Statistical Tests**. Figure 2 shows the flowchart of the algorithm.

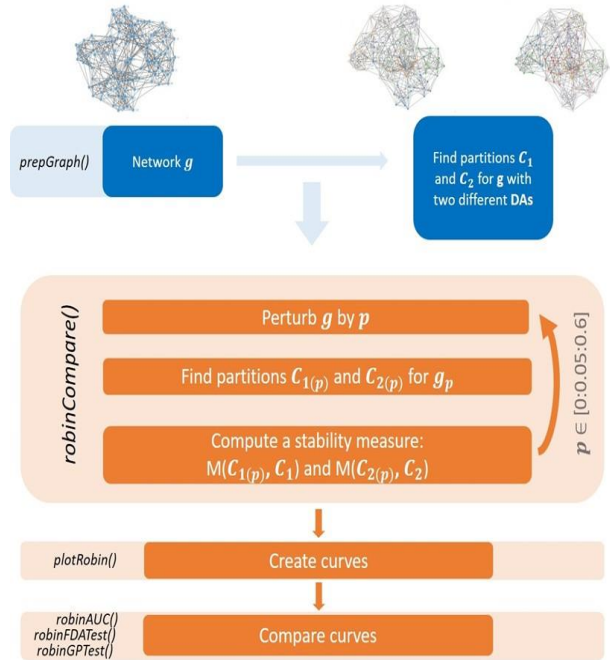

**Figure 3:** Flowchart summarizing the procedure to compare two different community detection algorithms.

This procedure allows the filtering of the detection algorithms according to their performance. Moreover, the selected ones can be compared using the second workflow. The second workflow helps to choose among different community detection algorithms the one that better fits the network of interest, comparing their robustness two at a time. More precisely, the technique:

1. finds two partitions $C_1$ and $C_2$ inferred by two different algorithms on the network under study,
2. perturbs the network creating a new one,
3. retrieves two new partitions $C_{1(p)}$ and $C_{2(p)}$,
4. evaluates $M\left(C_{1(p)}, C_1\right)$ and $M\left(C_{2(p)}, C_2\right)$.

Steps 2) - 4) are repeated at different perturbation levels $p \in [0 : 0.05 : 0.6]$ to create two curves and then their similarity is tested. Figure 3 shows the flowchart of the algorithm.

**Perturbation strategy**

The perturbed network has been restricted to have the same number of vertices and edges as the original unperturbed network, therefore only the edges position changes. It is expected that, if a community structure is robust, it should be stable under small perturbations of the edges. This is because perturbing the network edges by a small amount will imply just a low percentage of nodes to be moved in different communities; on the other hand, perturbing a high percentage of the edges in the network will produce random clusters. Note that a null percentage of perturbation $p = 0$ will correspond to the original graph while a maximal perturbation level $p = 1$ will correspond to the random graph. Therefore, in **robin** the perturbation of a network preserves the degree distribution of the original network.

Two different procedures for the perturbation strategy are implemented, namely independent and dependent type. The independent strategy introduces a percentage $p$ of perturbation in the original graph at each iteration, for $p = 0, \ldots, p_{max}$. Whereas the dependent procedure introduces 5% of perturbation at each iteration on the previous perturbed graph, starting from the original network, until $p_{max}$ of the graph's edges will be perturbed. In the implementation of the perturbation strategy, we set up $p_{max} = 0.6$, because the structure of the network becomes random if we perturb more than 50% of the edges.

In particular, we noticed that the greater modification of the network structure happens for a perturbation level between 30% and 40% if a network is robust, while it happens at very low perturbation levels if the network is not robust.

We stress again that the M curve for a network with a strong structure grows up rapidly (perturbation level between 0% and 40%) then levels off when $50\% < p < 100\%$.

Moreover, the choice $p_{max} = 0.6$ reduces the computational time and shows more clearly the differences between the curves.

Varying the percentage of perturbation, many graphs are generated and compared by means of the stability measure chosen. For each perturbation level, we generated 10 perturbed graphs and calculated the stability measure. From each of these graphs, we generated 9 more by rewiring an additional 1% of the edges. Therefore, the procedure generates 100 graphs with the respective stability measures for each level of $p$ and gives as output the mean of the stability measure for every 10 graphs generated.

**Stability measure**

The procedure we implemented is based on four different stability measures:

- the Variation of Information (VI) proposed by Meilă (2007),
- the Normalized Mutual Information (NMI) measure proposed by Danon et al. (2005),
- the split-join distance of van Dongen (2000),
- the Adjusted Rand Index (ARI) by Hubert and Arabie (1985).

VI measures the amount of information lost and gained in changing from one cluster to another, while split-join distance calculates the number of nodes that have to be exchanged to transform any of the two clusterings into the other; but for both of them low values represent more similar clusters and high values represent more different clusters. On the contrary, NMI and ARI are similarity measures therefore lower values identify more different clusters and higher values more similar ones. To make all the measures comparable, we considered the 1-1 transformation for the NMI and the ARI since they vary between [0, 1] as:

$$f(X) = 1 - X$$

Only two of the four proposed stability measures, i.e. split-join and VI, are distances. Moreover, they differ in their dependency on the number of clusters K: while the VI distance grows logarithmically with K, the split-join metric grows with K toward the upper bound of 1. To make the four different stability measures comparable, we normalized VI and split-join between 0 and 1 (i.e. we divided the VI and the split-join by their maximum, respectively $\log(n)$ and $2n$, where n indicates the number of vertices in the graph).

**Statistical Tests**

**robin** allows different multiple statistical tests to check the differences between the real and the random curve or between the curves built from two different detection algorithms. The variation of $p$ from 0 to 0.6 induces an intrinsic order to the data structure as in temporal data. This lets $p$ assume the same role as a time point in a temporal process and as a consequence, we can use any suitable time series approach to compare our curves. In the following, we describe the use of two such approaches.

The first is a test based on the Gaussian Process regression (GP) described in Kalaitzis and Lawrence (2011b). In this paper the authors use GP to compare treatment and control profiles in biological time-course experiments. The main idea is to test if two time series represent the same or two different temporal processes. A gaussian process is a collection of random variables, any finite number of which have a joint Gaussian distribution and is completely specified by its mean function and its covariance function, see e.g. Rasmussen and Williams (2006). Given the mean function $m(x)$ and the covariance function $k(x, x^J)$ of a real process $f(x)$ then we can write the GP as

$$f(x) \sim \mathsf{GP}(m(x), k(x, x^J)). \qquad (2)$$

The random variables $\mathbf{f} = (f(X_1), \ldots, f(X_n))^T$ represent the value of the function $f(x)$ at time locations $(X_i)_{i=1,\ldots,n}$, being $f(x)$ the true trajectory/profile of the gene. Assuming $f(x) = \Phi(x)^T \mathbf{w}$, where $\Phi(x)$ are projection basis functions, with prior $\mathbf{w} \sim N(\mathbf{0}, \sigma_{\mathbf{w}}^2 \mathbf{I})$, we have

$$m(x) = \Phi(x)^T E[\mathbf{w}] = 0, \quad k(x, x^J) = \sigma_{\mathbf{w}}^2 \Phi(x)^T \Phi(x) \tag{3}$$

$$f(x) \sim GP(0, \sigma_{\mathbf{w}}^2 \Phi(x)^T \Phi(x)). \tag{4}$$

Since observations are noisy, i.e. $\mathbf{y} = \mathbf{\Phi w} + \boldsymbol{\varepsilon}$, with $\mathbf{\Phi} = (\Phi(X_1)^T, \ldots, \Phi(X_n)^T)$, assuming that the noise $\boldsymbol{\varepsilon} \sim N(\mathbf{0}, \sigma_n^2 \mathbf{I})$ and using Eq. (3), the marginal likelihood becomes

$$p(\mathbf{y}|\mathbf{x}) = \frac{1}{(2\pi)^{\frac{n}{2}} \cdot |\mathbf{K_y}|} \exp\left(-\frac{1}{2} \mathbf{y}^t \mathbf{K_y}^{-1} \mathbf{y}\right), \tag{5}$$

with $\mathbf{K_y} = \sigma_{\mathbf{w}}^2 \mathbf{\Phi\Phi}^T + \sigma_n^2 \mathbf{I}$.

In this framework, the hypothesis testing problem over the perturbation interval $[0, p_{max}]$ can be reformulated as:

$$H_0 : \log_2 \frac{M(x)}{M_2(x)} \sim GP\left(0, k(x, x^J)\right) \quad \text{against} \quad H_1 : \log_2 \frac{M(x)}{M_2(x)} \sim GP\left(m(x), k(x, x^J)\right), \tag{6}$$

where $x$ represents the perturbation level. To compare the two curves, **robin** used the Bayes Factor (BF), that is approximated with a log-ratio of marginal likelihoods of two GPs, each one representing the hypothesis of differential (the profile has a significant underlying signal) and non-differential expression (there is no underlying signal in the profile, just random noise).

The second test implemented is based on the Interval Testing Procedure (ITP) described in Pini and Vantini (2016). The ITP provides an interval-wise non-parametric functional testing and is not only able to assess the equality in distribution between functions, but also to underline specific differences. Indeed, users can see where are localized the differences between the two curves. The Interval Testing Procedure is based on:

1. Basis Expansion: functional data are projected on a functional basis (i.e. Fourier or B-splines expansion);
2. Interval-Wise Testing: statistical tests are performed on each interval of basis coefficients;
3. Multiple Correction: for each component of the basis expansion, an adjusted p-value is computed from the p-values of the tests performed in the previous step.

In summary, GP provides a global answer on the dissimilarity of the two M curves while ITP points out local changes between such curves. As a rule of the thumb, we suggest initially using GP to flag a difference and then ITP to understand at which level of perturbation such a difference is locally significant.

We also provide a global method to quantify the differences between the curves when they are very close. This is based on the calculation of the area under the curves with a spline approach.

## Package structure

### Installation

Once in the R environment, it is possible to install and load **robin** package by using the standard installation function which retrieves the package from the CRAN repository and takes care of its dependencies, as follow:

```
install.packages("robin")
```

The **robin** package includes as dependencies **igraph** (Csardi and Nepusz, 2019), **networkD3** (C. Gandrud and Yetman, 2017), **ggplot2** (Wickham, 2019), **gridExtra** (Auguie, 2017), **fdatest** (Pini and Vantini, 2015) , **gprege** (Kalaitzis and Lawrence, 2011a) and **DescTools** (Signorell and mult. al., 2019) packages. All, except **gprege** which is a Bioconductor package, are automatically loaded with the command:

```
library(robin).
```

To install **gprege** package, start R and enter:

```
if (!requireNamespace("BiocManager", quietly = TRUE))
    install.packages("BiocManager")
BiocManager::install("gprege")
```

### Data import and visualization

**robin** is a user-friendly software providing some additional functions for data import and visualization, such as `prepGraph`, `plotGraph` and `plotComm`. The function `prepGraph`, required by the procedure, reads and simplifies undirected graphs removing loops and multiple edges. The available input graphs formats are: "edgelist", "pajek", "ncol", "lgl", "graphml", "dimacs", "graphdb", "gml", "dl" and an igraph object. The function `plotGraph`, with the aid of the **network3D** package, starting from an igraph object loaded with `prepGraph`, shows an interactive 3D graphical representation of the network, useful to visualize the network of interest before the analysis. Furthermore, the function `plotComm` helps to plot a graph with colourful nodes that simplifies the visualization of the detected communities, given the membership of the communities.

### Procedures

**robin**, embeds all the community detection algorithms present in **igraph**:

- `cluster_edge_betweenness`: it uses the concept of edge betweenness score that measures the number of shortest paths passing through the edge. It is based on an iterative procedure that calculates the betweenness of edges and removes those edges with the highest betweenness, getting a hierarchical map of the network (Newman and Girvan, 2004). The connected components are communities.
- `cluster_fast_greedy`: it is based on a greedy optimisation of the modularity and uses a hierarchical agglomerative approach for detecting community detection structure (Clauset et al., 2005).
- `cluster_infomap`: it is based on information-theoretic principles. It finds the community detection structure that minimizes the description length for a random walk on the graph given by the map equation over possible network partitions (Rosvall and Bergstrom, 2008).
- `cluster_leading_eigen`: it is based on a maximization process written in terms of the eigenvalues and eigenvectors of the modularity matrix, instead of the graph Laplacian commonly used in graph partitioning calculations (Newman, 2006).
- `cluster_louvain`: it implements the multi-level modularity optimization algorithm. The algorithm starts creating small communities looking at local modularity and then in an iterative procedure builds networks whose nodes are the communities. The process stops when maximum modularity is obtained and a hierarchy of communities is produced (Blondel et al., 2008).
- `cluster_label_prop`: it works by propagating labels through the network. Nodes are initialized with unique labels and updated to the labels of the maximum numbers of their neighbours iteratively. The algorithm reaches convergence when none of the nodes needs to update its label anymore (U. N. Raghavan, 2007).
- `cluster_spinglass`: it is based on the Potts model. In this model, each node can be in a spin state and the edges specify the pairs of nodes that stay in the same spin state. The model is simulated several times and communities are composed of nodes in the same state (Reichardt and Bornholdt, 2006) .
- `cluster_walktrap`: it defines a measure of similarity between nodes in terms of random walks and uses an agglomerative method to group iteratively nodes into communities. The basic idea is that short random walks on a graph tend to stay in the same community (Pons and Latapy, 2005).

**robin** gives the possibility to input a custom external function to detect the communities. The user can provide his own function as value of the parameter `FUN` in both analyses, implemented into the functions `robinRobust` and `robinCompare`. These two functions create the internal process for perturbation and measurement of communities stability. In particular `robinRobust` tests the robustness of a chosen detection algorithm and `robinCompare` is specifically designed to compare two different detection algorithms. The option `measure` in the `robinRobust` and `robinCompare` functions provides the flexibility to choose between the four different measures listed in subsection **Stability measure**.

**robin** offers two choices for the null model to set up for `robinRobust`:

- external building according to users' preferences, then the null graph is passed as a variable,
- generation by using the function `random`.

The function `random` creates a random graph with the same degree distribution of the original graph, but with completely random edges, by using the `rewire` function of the **igraph** package with the

`keeping_degseq` option that preserves the degree distribution of the original network. The function `rewire` randomly assigns a number of edges between vertices with the given degree distribution. Note that **robin** assumes undirected graphs without loops and multiple edges which are directly created, from any input graph, by the function `prepGraph`

**Construction of curves**

The `plotRobin` function allows the user to generate two curves based on the computation of the chosen stability measures.

When `plotRobin` is used considering as input parameters the output of `robinRobust`, i.e. the first step of the overall procedure, the first curve represents the measure between the partition of the original unperturbed graph and the partition of each perturbed graph (blue curve in Figure 4-Left panel), and the second curve is obtained in the same way but considering as the original graph the random graph (red curve in Figure 4-Left panel). The comparison between the two curves turns the question about the significance of the retrieved community structure into the study of the robustness of the recovered partition against perturbation.

When `plotRobin` is used considering as input parameters the output of `robinCompare`, i.e. the second step of the overall procedure, it generates a plot that depicts two curves, one for each clustering algorithm. In the right panel of Figure 4 each curve is obtained computing the measure between the partition of the original unperturbed graph with the partition of each perturbed graph, where the partition method is either Louvain (blue curve) or Fast Greedy (red curve).

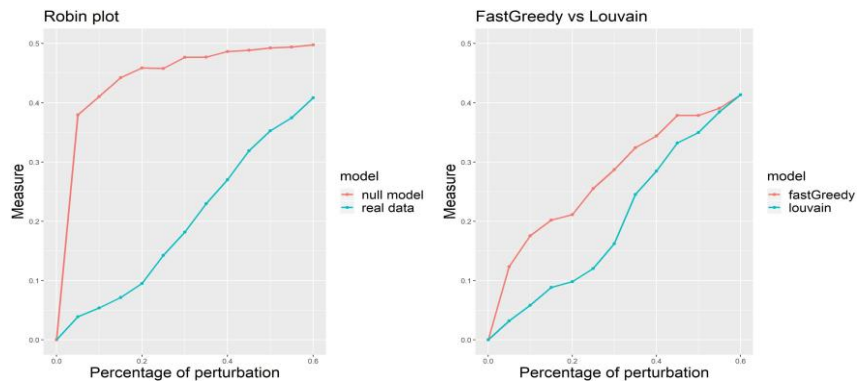

**Figure 4:** Curves of the null model and the real data generated by the VI stability measure and the Louvain detection algorithm on the American College Football network (Left panel). Curves of the Louvain and Fast greedy algorithm generated by the VI stability measure on the American College Football network (Right panel) (Girvan and Newman, 2002).

**Testing**

The GP test is implemented in `robinGPTest` and uses the R package **gprege** (Kalaitzis and Lawrence, 2011a). The ITP test is implemented in `robinFDATest` and uses the R package **fdatest** (Pini and Vantini, 2015). The area under the curves is calculated by the function `robinAUC` and relies on the **DescTools** package. Figure 5 shows the curves for the comparison of Louvain and Fast greedy algorithms' performance generated by the VI stability measure using the Interval Testing Procedure on the American College Football network (left panel) (Girvan and Newman, 2002) and corresponding adjusted p-values (right panel).

All the functions implemented in **robin** are summarized in Table 2.

**Computational time**

The time complexity of the proposed strategy, more precisely of the `robinRobust` function, is evaluated as follows. Generating a rewired network with $N$ nodes and $M$ edges consumes $O(N + M)$ time, for both the real and the null model. For each network we detect the communities, using any community detection algorithm present in **igraph** or any custom external algorithm inserted by the user, and calculate a stability measure. Let $D$ be the time complexity associated with the community detection algorithm chosen. The overall procedure is iterated $k = 100$ times for each percentage $p$ of the $n_p = 12$ perturbation levels ($p \in [0, p_{max}], p_{max} = 0.6$). In total the proposed procedure requires $O(D + (((N + M + D) * k) * n_p))$ time both for the real and the null model.

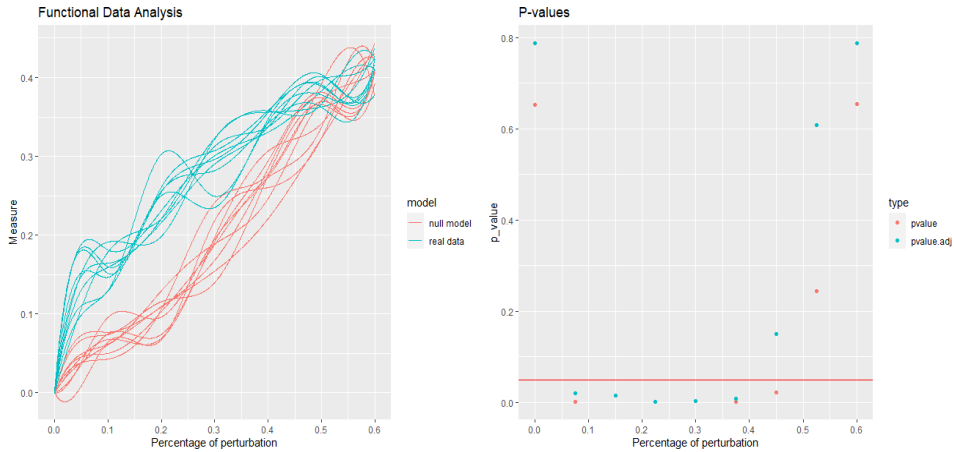

**Figure 5:** Curves of the Louvain and Fast greedy algorithm generated by the VI stability measure using the Interval Testing Procedure on the American College Football network (Left panel). Corresponding p-values and adjusted p-values for all the intervals with the horizontal red line on to the critical value 0.05 (Right panel)
.

In Table 1 we show the computational time evaluated on a computer with an Intel 4 GHz i7-4790K processor and 24GB of memory. In this example, we used *Louvain* as a detection algorithm on the *LFR* benchmark graphs (Lancichinetti et al., 2008). The time complexity could be mitigated using parallel computing but this not yet implemented.

**Table 1:** Computational time

| NODES | EDGES | TIME(SECS) |
|---|---|---|
| 100 | 500 | 2.1 |
| 1000 | 9267 | 36.1 |
| 10000 | 100024 | 361.8 |
| 100000 | 994053 | 9411.6 |
| 1000000 | 8105913 | 110981.5 |

## Example test: the American College football network

**robin** includes the *American College football* benchmark dataset as an analysis example that is freely available at http://www-personal.umich.edu/~mejn/netdata/. The dataset contains the network of United State football games between Division I colleges during the 2000 season (Girvan and Newman, 2002). It is a network of 115 vertices that represent teams (identified by their college names) and 613 edges that represent regular-season games between the two teams they connect. The graph has the ground truth, where each node has a value that indicates to which of the 12 conferences it belongs, and this offers a good opportunity to test the ability of **robin** to validate the community robustness. It is known that each conference contains around 8-12 teams, the games are more frequent between members of the same conference than between members of different conferences, they are on average seven between teams of the same conference and four between different ones. We applied all the methods listed in subsection **Procedures** to this network, choosing as measure the VI metric.

```
library(robin)

my_network <- system.file("example/football.gml", package="robin")
graph <- prepGraph(file=my_network, file.format="gml")

attributes<-vertex_attr(graph, index = V(graph))
real<-attributes$value
real<-as_membership(real)
```

Table 2: Summary of the functions implemented in **robin**.

| FUNCTION | DESCRIPTION |
| --- | --- |
| **Import/Manipulation** | |
| prepGraph | Management and preprocessing of input graph |
| random | Building of null model |
| **Analysis** | |
| robinRobust | Comparison of a community detection method versus random perturbations of the original graph |
| robinCompare | Comparison of two different community detection methods |
| **Visualization** | |
| methodCommunity | Detection of the community structure |
| membershipCommunities | Detection of the membership vector of the community structure |
| plotGraph | Graphical interactive representation of the network |
| plotComm | Graphical interactive representation of the network and its communities |
| plotRobin | Plots of the two curves |
| **Test** | |
| robinGPTest | GP test and evaluation of the Bayes factor |
| robinFDATest | ITP test and evaluation of the adjusted p-values |
| robinAUC | Evaluation of the area under the curve |

```
set.seed(10)
members_In<-membershipCommunities(graph=graph, method=DA)
VI_In <- compare(real, members_In, method="vi")
```

Note that the variable *DA* refers to the detection algorithms present in **igraph** and can assume the following values: *fastGreedy*, *infomap*, *walktrap*, *edgeBetweenness*, *spinglass*, *leadingEigen*, *labelProp*, *louvain*. The function compare is contained in the package **igraph** and permits the assessment of the distance between two community structures according to the chosen method.

Table 3 summarizes the VI results calculated between the real communities and the ones that the detection algorithms created.

It is possible to observe that the best performance is offered by Infomap, having the lowest VI value, followed by Spinglass. Louvain, Propagating Labels, Walktrap and Edge betweenness have a similar intermediate VI value, while the worst performance is given by Fast greedy and Leading eigenvector. Then, we used **robin** to check if the results are confirmed looking at the VI curves and the results of the testing procedure for the second workflow, i.e. the one comparing two detection algorithms, considering Infomap versus all the others.

**Table 3:** VI measure between different methods and ground-truth.

| METHODS | NORMALIZED VI |
|---|---|
| cluster_infomap | 0.0543488 |
| cluster_spinglass | 0.0625997 |
| cluster_louvain | 0.0758894 |
| cluster_label_prop | 0.0763273 |
| cluster_walktrap | 0.0780843 |
| cluster_edge_betweenness | 0.0833228 |
| cluster_fast_greedy | 0.1853927 |
| cluster_leading_eigen | 0.1956331 |

```
comp <- robinCompare(graph=graph, method1=DA1,
            method2=DA2, measure="vi", type="independent")

plotRobin(graph=graph, model1=comp$Mean1, model2=comp$Mean2,
            measure="vi")
```

Figure 6 shows the results we obtained. If we focus on the perturbation interval [0, 0.3], it is possible to note the similar behaviour between the curves representing Infomap/Spinglass, Infomap/Louvain, Infomap/Propagating Labels, Infomap/Walktrap and Infomap/Edge betweenness, with a closer distance between Infomap/Spinglass. On the contrary, the curves Infomap/Fast greedy and Infomap/Leading eigenvector have an opposite behaviour, building almost an ellipse. This confirms what displayed in Table 3.

In our overall procedure, we explored two different ways of generating a null model, namely the Configuration Model (CM) and the *dk*-series model.

```
graphRandomCM<- random(graph=graph)
graphRandomDK<- prepGraph(file="dk2.1_footballEdgelist.txt",
                    file.format = "edgelist")

plotGraph(graph)
plotGraph(graphRandomCM)
plotGraph(graphRandomDK)
```

The different structures provided by the real data network, CM and *dk*-series with $d = 2.1$ are shown in Figure 7.

The CM generates a random graph with the same degree sequence of the original one but with a randomized group structure. Our experiments show that CM is not a good null model when using Propagating Labels and Infomap as community extraction methods (Figure 8). In fact, when the modularity is low, these two algorithms tend to assign all the nodes to the same community, hence resulting in a flat stability measure curve. We launched the function robinRobust to assess the robustness of each detection algorithm.

```
proc_CM <- robinRobust(graph=graph, graphRandom=graphRandomCM,
                    measure="vi", method=DA, type="independent")

plotRobin(graph=graph, model1=proc_CM$Mean, model2=proc_CM$MeanRandom,
            measure="vi")
```

The *dk*-series model generates a random graph preserving the global organization of the original network at various increasing levels of details chosen by the user via the setting of the parameter *d*. In particular, we chose the *dk* random graph with d=2.1.

```
proc_DK <- robinRobust(graph=graph, graphRandom=graphRandomDK,
                    measure="vi", method="fastGreedy", type="independent")
```

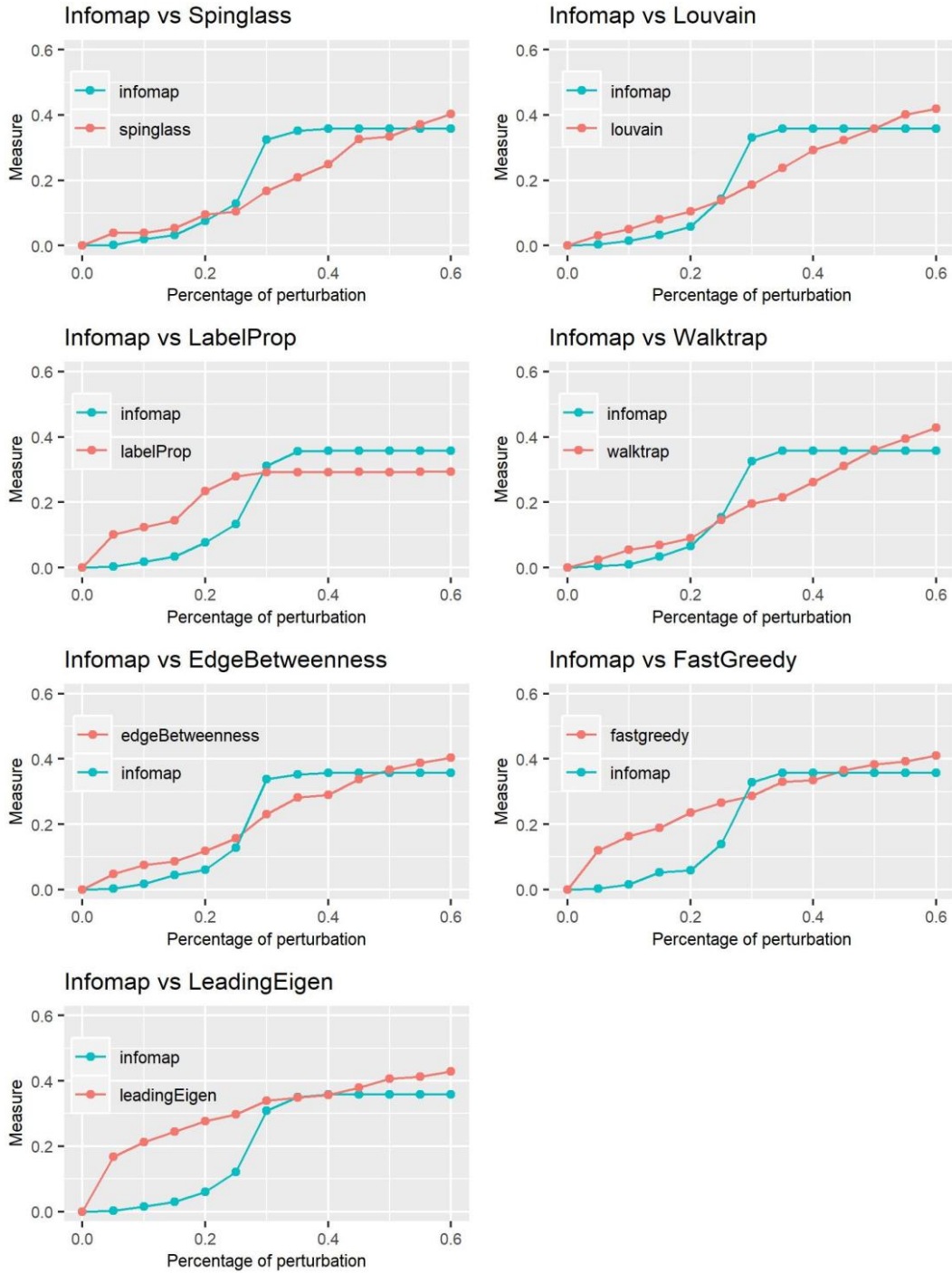

**Figure 6:** Plot of the VI curves of Infomap against all other methods.

```
plotRobin(graph=graph, model1=proc_DK$Mean, model2=proc_DK$MeanRandom,
          measure="vi")
```

Figure 9 shows the stability measure curves of each detection algorithm compared to dk 2.1 null model. For all the methods, the two curves are very close due to the capability of the null model to preserve a structure similar to the real network and visually confirm the results in Table 3.

Moreover, for the *dk*-series model we tested the differences between the two curves using the GP methodology implemented in the function in `robinGPTest`.

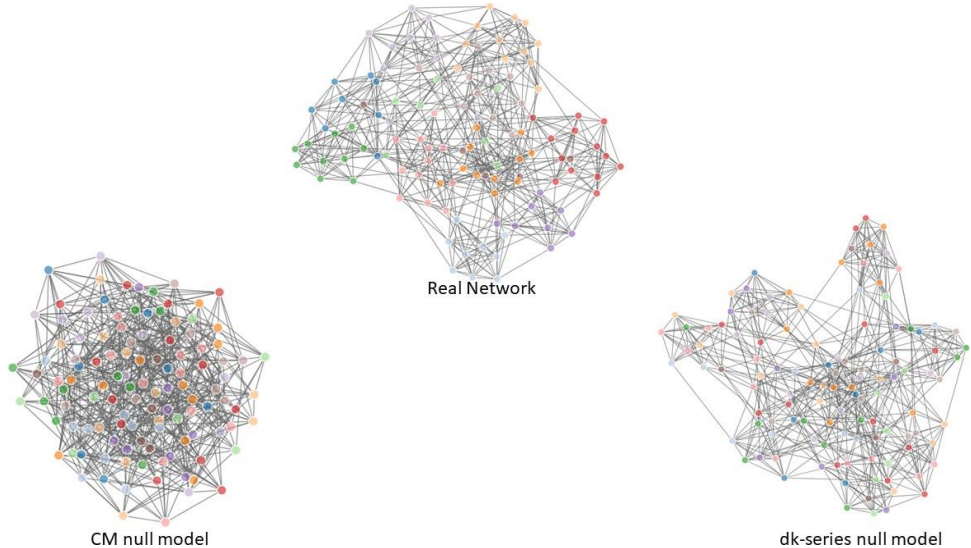

**Figure 7:** Graph of the real data (Upper panel); graph of the CM null model (Left - lower panel); graph of the *dk*-series null model with $d = 2.1$ (Right - lower panel).

```
BFdk <- robinGPTest(model1=proc_DK$Mean, model2=proc_DK$MeanRandom)
```

Results are shown in Table 4 and are in agreement with those shown in Table 3.

**Table 4:** Bayes Factor and AUC ratio for *dk*-series with $d = 2.1$

| METHODS | BAYES FACTOR | AUC |
|---|---|---|
| cluster_infomap | 53.66558912 | 1.133480375 |
| cluster_spinglass | 40.55115149 | 1.249467983 |
| cluster_louvain | 31.52079568 | 1.208205961 |
| cluster_label_prop | 102.2408283 | 0.85227387 |
| cluster_walktrap | 30.73857577 | 1.203845609 |
| cluster_edge_betweenness | 31.09826284 | 1.193820919 |
| cluster_fast_greedy | 0.001932996 | 1.017251935 |
| cluster_leading_eigen | 8.474086292 | 1.042145119 |

Fastgreedy clearly fails in recovery the communities, LeadingEigen has stronger evidence but too weak when compared to the other methods. Louvain, Walktrap and EdgeBetweenness have the same strong evidence followed by Spinglass and Infomap. LabelProp shows the strongest evidence but the result is obviously influenced by the swap between the two curves when the perturbation is greater than 20%, underlying a worse performance of the algorithm. The same swap can be noted for Infomap at 35% perturbation level, but with a less difference between the two curves. This is confirmed by the fact that the ratios between the AUC of the real null model curve and the AUC of the real network are close to 1.

```
AUC <- robinAUC(graph=graph, model1=proc_DK$Mean,
                model2=proc_DK$MeanRandom, measure="vi")
AUCdkratio <- AUC$area2/AUC$area1
```

Also note that LabelProp originate the paradox that AUC of the real model curve exceeds the AUC of the null network, despite the hypothesis testing result is positive. Hence it is always the case to look at the plots and AUC ratios.

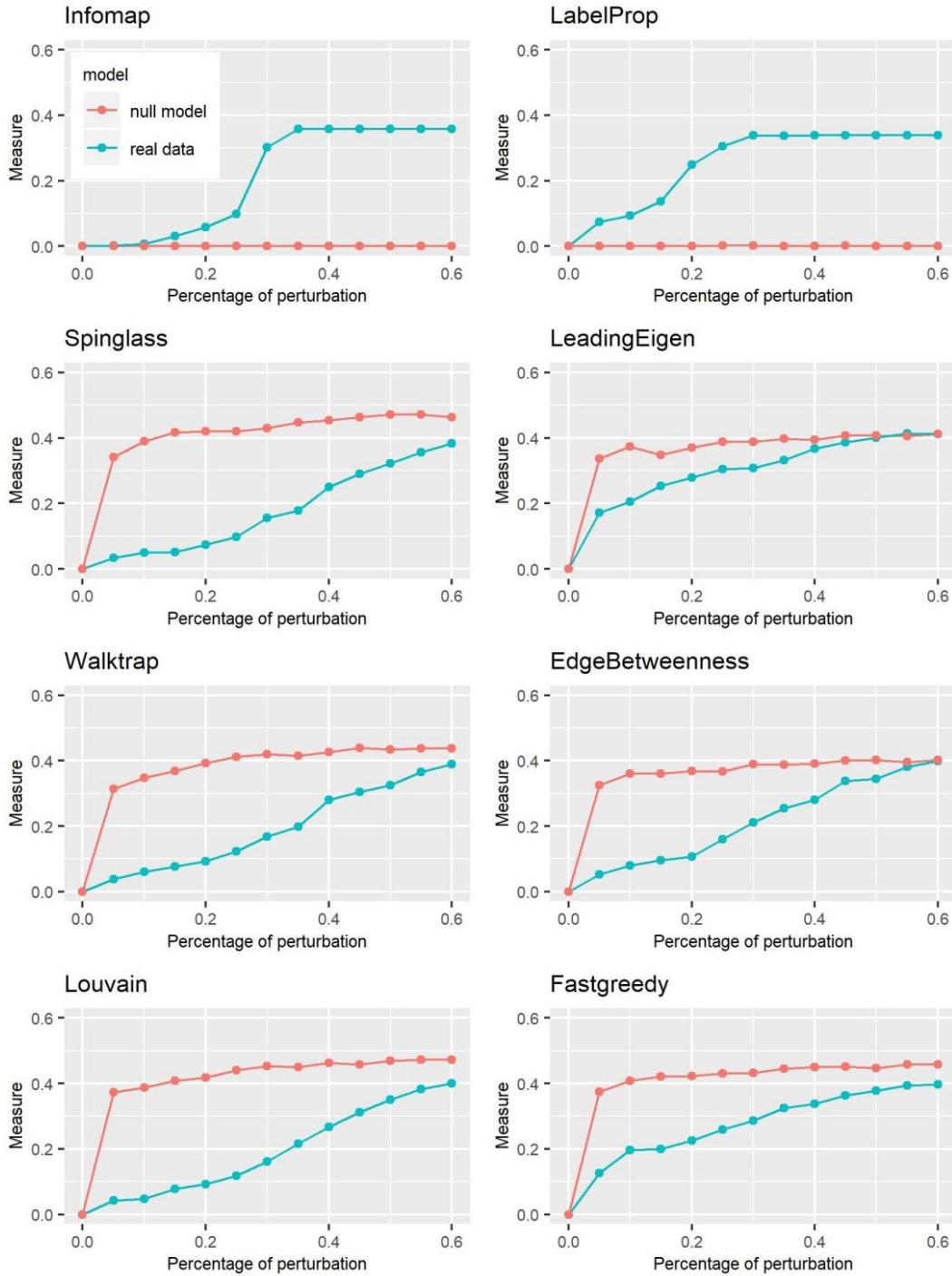

**Figure 8:** Plot of the VI curves of the CM null model and all the algorithms implemented.

## Conclusion

In this paper, we presented **robin**, an R/CRAN package to assess the robustness of the community structure of a network found by one or more detection methods, providing an indication about their reliability. The procedure implemented is useful to compare different community detection algorithms and choose the one that best fits the network of interest. More precisely, **robin** initially detects if the community structure found by some algorithms is statistically significant, then it compares the different selected detection algorithms on the same network. **robin** uses analysis tools set up for functional data analysis, such as *GP* regression and *ITP*. The core functions of the package are `robinRobust` and `robinCompare`, that build the stability measure curves for the null model and the network understudy for a fixed detection algorithm and the stability measure for the network understudy for two detection algorithms, respectively; and `robinGPTest` and `robinFDATest`, that implement the GP test and the ITP test. We illustrated the usage of the package on a benchmark dataset. The package is available on CRAN at https://CRAN.R-project.org/package=robin.

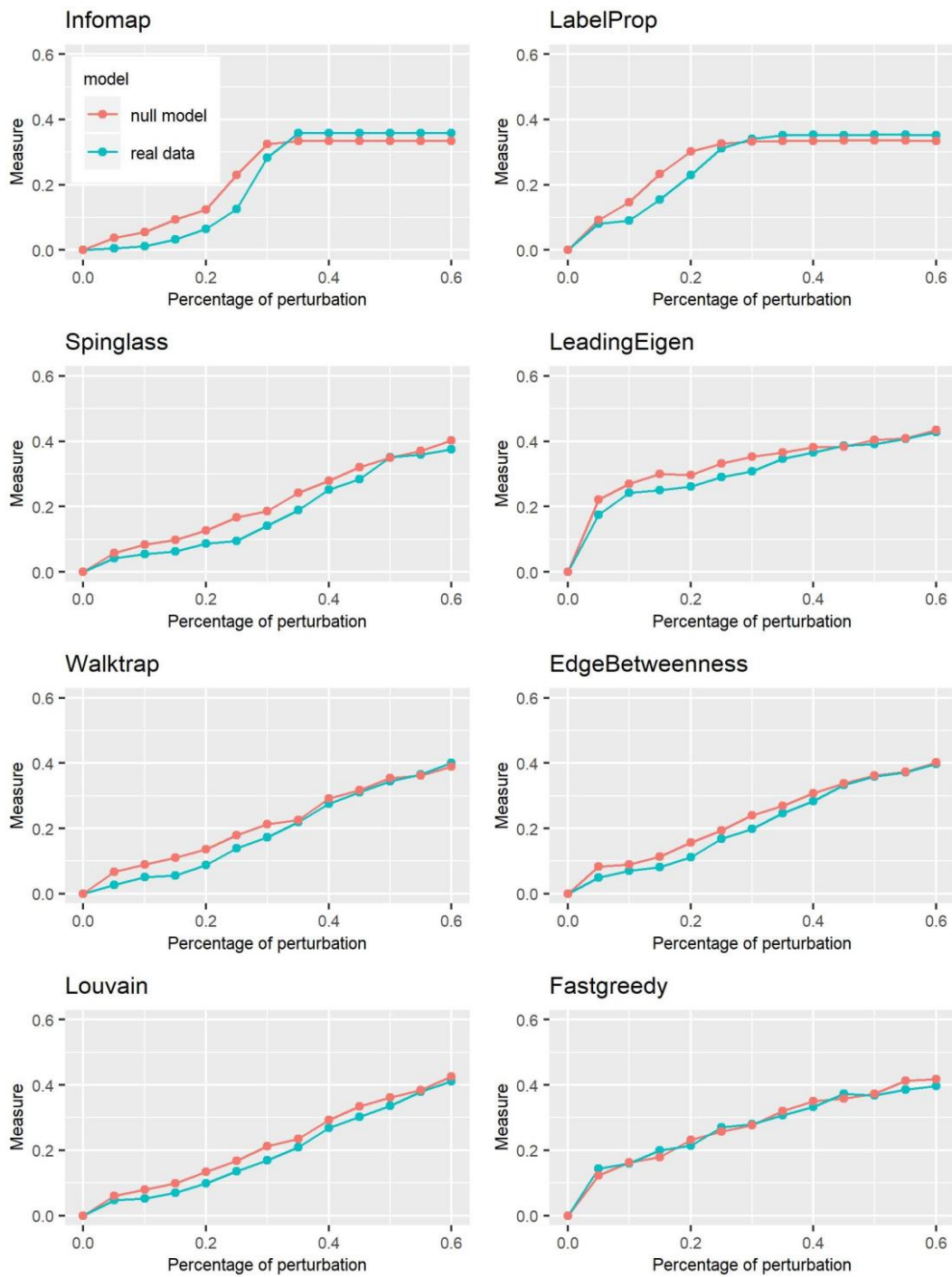

**Figure 9:** Plot of the VI curves of the *dk*-series null model with $d = 2.1$ and all the algorithms implemented.

## Computational details

The results in this paper were obtained using R 3.6.1 with the packages **igraph** version 1.2.4.2, **networkD3** version 0.4, **ggplot2** version 3.2.1, **gridExtra** version 2.3, **fdatest** version 2.1, **gprege** version 1.30.0 and **DescTools** version 0.99.31. R itself and all packages used are available from the Comprehensive R Archive Network (CRAN) at https://CRAN.R-project.org/.

## Acknowledgments

This work was supported by the project Piattaforma Tecnologica ADVISE - Regione Campania.

## Contributions

V.P. implemented the software and analysed its properties. D.R. supported V.P. in R package implementation. A.C., L.C. and I.D.F. conceived the work, contributed to the development and implementation of the concept, discussed and analysed the results. A.C., L.C., I.D.F. and V.P. wrote the manuscript.

*Valeria Policastro (package author and creator)*
*Dipartimento Scienze e Tecnologie Ambientali, Biologiche e Farmaceutiche*
*Universitá degli Studi della Campania "Luigi Vanvitelli"*
*Via Vivaldi, 43 81100 Caserta, Italia*
*and*
*Istituto per le Applicazioni del Calcolo "M. Picone" - sede di Napoli*
*Consiglio Nazionale delle Ricerche*
*via Pietro Castellino 111*
*80131 Napoli, Italy*
valeria.policastro@unicampania.it

*Dario Righelli*
*Dipartimento di*
*Statistica Universitá di*
*Padova*
*Via Cesare Battisti, 241 35121 Padova, Italia*

*and*
*Istituto per le Applicazioni del Calcolo "M. Picone" - sede di Napoli*
*Consiglio Nazionale delle Ricerche*
*via Pietro Castellino 111*
*80131 Napoli, Italy*
d.righelli@na.iac.cnr.it

*Annamaria Carissimo (corresponding author)*
*Istituto per le Applicazioni del Calcolo "M. Picone" - sede di Napoli*
*Consiglio Nazionale delle Ricerche*
*via Pietro Castellino 111*
*80131 Napoli, Italy*
a.carissimo@iac.cnr.it

*Luisa Cutillo (corresponding author)*
*School of Mathematics*
*University of Leeds*
*Leeds*
*LS2 9JT, United Kingdom*
*and*
*Dipartimento di Studi Aziendali e Quantitativi*
*Universitá degli Studi di Napoli "Parthenope'*
*Via Generale Parisi, 13*
*80132, Napoli, Italia*
l.cutillo@leeds.ac.uk

*Italia De Feis (corresponding author)*
*Istituto per le Applicazioni del Calcolo "M. Picone" - sede di Napoli*
*Consiglio Nazionale delle Ricerche*
*via Pietro Castellino 111*
*80131 Napoli, Italy*
i.defeis@iac.cnr.it